\documentclass{article}
\usepackage{ijcai25}

\usepackage{times}
\usepackage{soul}
\usepackage{url}
\usepackage[hidelinks]{hyperref}
\usepackage[utf8]{inputenc}
\usepackage[small]{caption}
\usepackage{graphicx}
\usepackage{amsmath}
\usepackage{amsthm}
\usepackage{booktabs}
\usepackage{algorithm}
\usepackage{algorithmic}
\usepackage[switch]{lineno}
\usepackage{multirow}
\usepackage{amsfonts}

\title{M3ANet: Multi-scale and Multi-Modal Alignment Network for Brain-Assisted Target Speaker Extraction}

\author{
Cunhang Fan$^1$
\and
Ying Chen$^1$\and
Jian Zhou$^1$\and
Zexu Pan$^2$\and
Jingjing Zhang$^1$\and
Youdian Gao$^1$\and
Xiaoke Yang$^1$\and
Zhengqi Wen$^3$\and
Zhao Lv$^{1,}$\footnote{Corresponding author}\\
\affiliations
$^1$School of Computer Science and Technology, Anhui University, China\\
$^2$Alibaba group, Singapore\\
$^3$Department of Automation, Tsinghua University Beijing National Research Center for lnformation Science and Technology, Tsinghua University, China\\
\emails
\{cunhang.fan, jzhou, kjlz\}@ahu.edu.cn,
\{e23201035, e22201067, e23301143, e22201014\}@stu.ahu.edu.cn,
xu325504@hotmail.com,
zqwen@tsinghua.edu.cn
}

\begin{document}

\maketitle

\begin{abstract}
    The brain-assisted target speaker extraction (TSE) aims to extract the attended speech from mixed speech by utilizing the brain neural activities, for example Electroencephalography (EEG). However, existing models overlook the issue of temporal misalignment between speech and EEG modalities, which hampers TSE performance. In addition, the speech encoder in current models typically uses basic temporal operations (e.g., one-dimensional convolution), which are unable to effectively extract target speaker information. To address these issues, this paper proposes a multi-scale and multi-modal alignment network (M3ANet) for brain-assisted TSE. Specifically, to eliminate the temporal inconsistency between EEG and speech modalities, the modal alignment module that uses a contrastive learning strategy is applied to align the temporal features of both modalities. Additionally, to fully extract speech information, multi-scale convolutions with GroupMamba modules are used as the speech encoder, which scans speech features at each scale from different directions, enabling the model to capture deep sequence information. Experimental results on three publicly available datasets show that the proposed model outperforms current state-of-the-art methods across various evaluation metrics, highlighting the effectiveness of our proposed method. The source code is available at: https://github.com/fchest/M3ANet.
\end{abstract}

\section{Introduction}

The cocktail party problem~\cite{haykin2005cocktail} refers to the ability of humans to track a particular conversation in a multi-speaker social setting, known as \textit{selective auditory attention}. Target speaker extraction (TSE) is a fundamental task in speech signal processing that aims to extract the target speech using auxiliary cues, thus effectively addressing the cocktail party problem.

Recent advances in deep learning have significantly improved TSE performance~\cite{ge2020spex+,hao2024x}. Traditional TSE approaches often rely on pre-recorded speech from the target speaker as a cue~\cite{xu2020spex}, but this method requires prior knowledge of the target speaker. Other studies have explored alternative cues, such as spatial cues (e.g., the direction of the target speaker)~\cite{wang2024study,ge2022spex} or visual cues (e.g., facial video)~\cite{lin2023av,li2024audio}. However, in practical applications, visual cues can be hindered by obstacles in the environment, and spatial cues are inherently limited to multi-microphone setups~\cite{zmolikova2023neural}. These challenges, in turn, degrade the performance of TSE in complex, real-world environments.

Recent studies on auditory attention decoding (AAD) have revealed the correlation between brain neural activities and speech stimuli~\cite{yan2024darnet,ni2024dbpnet}, opening a new avenue for TSE techniques. Researchers are increasingly exploring how to leverage speech-related information embedded in Electroencephalography (EEG) to guide target speech extraction, thereby eliminating the need for prior speaker knowledge. Early methods focus on reconstructing speech envelope from EEG data and matching it with sound sources to identify target speech~\cite{han2019speaker,o2017neural}, but this approach is computationally intensive due to the need to separate all sources. The brain-informed speech separation (BISS) network~\cite{ceolini2020brain} innovatively integrates the reconstructed envelope into the TSE task, using it as an audio cue to avoid separating all sources. However, stimulus reconstruction requires a distinct network, which still entails considerable computational costs. More recent studies~\cite{qiu2023tf,pan2024neuroheed+} have further streamlined this process by directly using EEG signals as the reference cue, and these methods have since become the standard in the field.

However, EEG and speech data are simultaneously fed into the models during training, yet they are not perfectly synchronized. This is because sound stimuli must first pass through the ear and auditory pathways before reaching the auditory cortex, where EEG signals are generated~\cite{litovsky2015development}. Previous works~\cite{de2024neurospex,Zhang2023basen} only focus on the effectiveness of modality fusion but neglect the temporal alignment between speech and EEG, which increases the difficulty of efficient multi-modal fusion and negatively impacts overall performance. Additionally, convolutional neural networks (CNNs) used in most TSE speech encoders~\cite{hao2023typing} often fail to effectively capture the latent features of mixed speech, leading to reduced model performance. While transformer-based encoders~\cite{fan2025improved,heo2024centroid} offer more power, they tend to introduce significant computational complexity.

In this paper, we propose a multi-scale and multi-modal alignment network (M3ANet) for brain-assisted TSE, which facilitates temporal alignment between EEG and speech, while extracting more comprehensive and deeper speech features from both the breadth and depth. To address the misalignment caused by the temporal asynchrony between EEG and speech, this paper introduces a modal alignment module into the brain-assisted TSE task for the first time. It employs a contrastive learning strategy to minimize the distance between EEG and speech features at the same time steps, thereby achieving precise synchronization of cross-modal data. Additionally, to fully capture the latent features in speech, the speech encoder first uses multi-scale convolutions to extract short- and long-term features, obtaining local- and global-level temporal representations. Next, GroupMamba (GM) modules are introduced to efficiently model the channel and feature dimensions of speech from different directions, capturing deeper features at each level with linear complexity.  
Experimental results show that the M3ANet model achieves state-of-the-art (SOTA) performance on the Cocktail Party, AVED, and MM-AAD datasets, with relative improvements of 8.2\%, 12.8\%, and 10.4\% in scale-invariant source to distortion ratio (SI-SDR), respectively, compared to the best current baselines.

The main contributions of this paper can be summarized as follows:
\begin{itemize}
    \item This paper proposes a new brain-assisted target speaker extraction model, introducing an innovative modal alignment module that leverages contrastive learning to align EEG and speech temporally.
    \item This paper proposes a multi-scale encoding approach combined with the GM module, which effectively extracts latent speech features in both breadth and depth.
    \item Experimental results on the Cocktail Party, AVED, and MM-AAD datasets show significant improvements over baseline models, demonstrating the effectiveness of our proposed model.
\end{itemize}

\section{Related Works}
Neuroscience research has revealed that humans possess selective auditory attention, wherein target speech signals are closely linked to neural activity in the brain. This makes it possible to utilize EEG signals as references for target speaker extraction. O’sullivan et al.~\shortcite{o2015attentional} proposed that auditory stimuli can be decoded from neural responses (i.e., EEG) through a process called stimulus reconstruction~\cite{vanthornhout2018speech}. BISS~\cite{ceolini2020brain} is a pioneering brain-assisted TSE model that jointly trains a stimulus reconstruction network and a speaker extraction network to estimate the target speech. BESD~\cite{hosseini2021speaker} is the first model to directly use EEG signals as references in this field, enabling training with a single extraction network and reducing computational complexity. BESD employs a speech encoder consisting of two convolutional layers. The improved version, U-BESD~\cite{hosseini2022end}, enhances feature extraction by introducing skip connections and a U-shaped structure in the encoder-decoder. However, the performance of these models remains limited. BASEN~\cite{Zhang2023basen}, building upon ConvTasNet~\cite{luo2019conv}, uses 8 layers of depthwise convolutions to better capture speech information and incorporates an effective multi-layer cross-attention fusion module, leading to substantial performance improvements. MSFNet~\cite{fan2024msfnet}, which employs a multi-scale parallel extraction network, applies convolutions at different scales to extract speech features more effectively and currently holds the SOTA performance on the Cocktail Party dataset. NeuroHeed~\cite{pan2024neuroheed}, based on the Spex~\cite{xu2020spex} architecture, introduces an online self-registration auditory cue network and is the first to consider real-time processing. 

However, these networks do not account for the inherent temporal discrepancies between EEG and speech, which limits further performance improvements. Additionally, the simple convolution operations in their speech encoders result in inadequate feature representation.

\section{Model Architecture}
\begin{figure*}[t] %
    \centering 
    \includegraphics[width=1\textwidth]{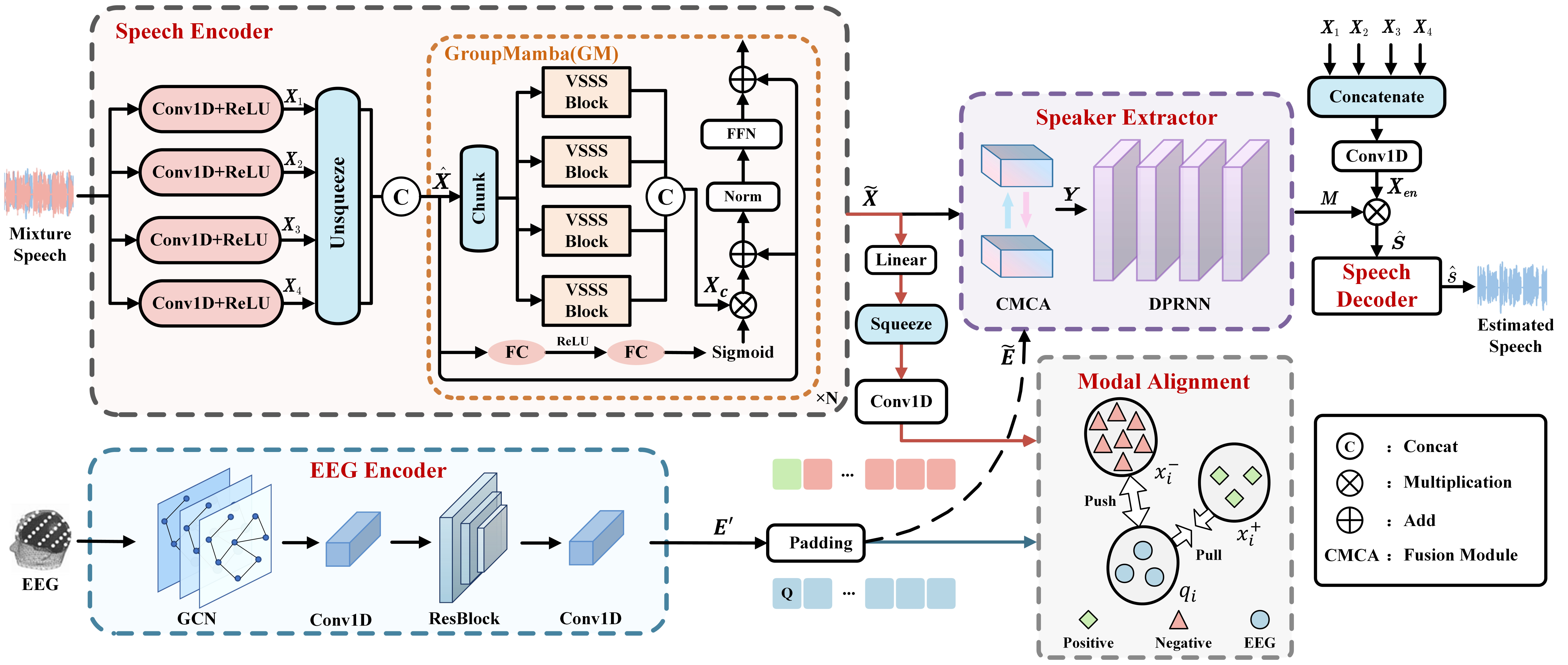}
    \caption{The overall architecture of the proposed method. The colored squares in the bottom right corner explain how the contrast learning strategy used in this paper defines query vectors and pairs of positive and negative samples. Each EEG segment is treated as a query vector; the temporally aligned speech segment serves as the positive sample, while the other segments within the same batch are considered negative samples. The CMCA is a convolutional multi-layer cross attention module.} 
    \label{fig:overall} 
\end{figure*}
The proposed model primarily consists of four components: the speech encoder, EEG encoder, modal alignment module, and speaker extractor. The system workflow is as shown in Figure~\ref{fig:overall}. The speech and EEG encoders extract feature representations of the input signals, which are then fed into the fusion module for effective information integration, while also being passed through the alignment module for modal alignment. The fused features are then passed through a speaker extractor to generate an estimated mask, which is multiplied by the encoded speech features to obtain the masked speech. Finally, the target speaker's speech is decoded through a 1D transposed convolution. Each part is described in detail below.

\subsection{EEG Encoder} 
The structure of the whole EEG encoder is shown in Figure~\ref{fig:overall}. In this paper, we use the same EEG encoder with graph convolutional network (GCN)~\cite{ren2024phase} as in~\cite{fan2024msfnet}. 
If the EEG data are recorded with N electrodes, it can be abstracted as a graph $G = (V,E)$ (Here, E denotes the set of edges, while elsewhere in the paper, it represents the EEG data), where $V$ is the set of $|V|=N$ channels, and $E\subset V\times V$ is the set of undirected connections between channels. 

Initially, a three-layer graph convolution operation is used on each EEG segment to generate graph-level embedding, and the chebyshev polynomial is used as the convolution kernel of GCN to simplify the laplacian matrix computation. Next, we normalize the data along the channel dimension using channel-wise normalization to achieve a more uniform feature distribution. The data then sequentially pass through a 1×1 Conv1D layer, followed by three ResBlocks, and are finally processed by another 1×1 Conv1D layer to produce the encoded EEG embedding vector $E^{'} \in \mathbb{R}^{B \times N_e \times T_e}$. The ResBlock consists of two convolution layers, each of them followed by a batch normalization (BN) layer and a PReLU activation function, with residual connections introduced after the second BN layer. Additionally, a MaxPooling layer is included to capture more representative temporal features. Now let $R$ denote the input features and $R^{'}$ the output of the first convolution module. The ResBlock can be represented as follows:
\begin{equation} 
    R^{'} = PReLU(BN(Conv1D(R,1,N_{in},N_{out})))
\end{equation}
\begin{equation}
    \resizebox{.91\linewidth}{!}{$
        \displaystyle
    X_{res} = PReLU(R + BN(Conv1D(R^{'},1,N_{out},N_{out})))
    $}
\end{equation}
where $N_{in}$, $N_{out}$ represent the number of channels.

\subsection{Speech Encoder}
\begin{figure}[h] 
    \centering 
    \includegraphics[width=0.45\textwidth]{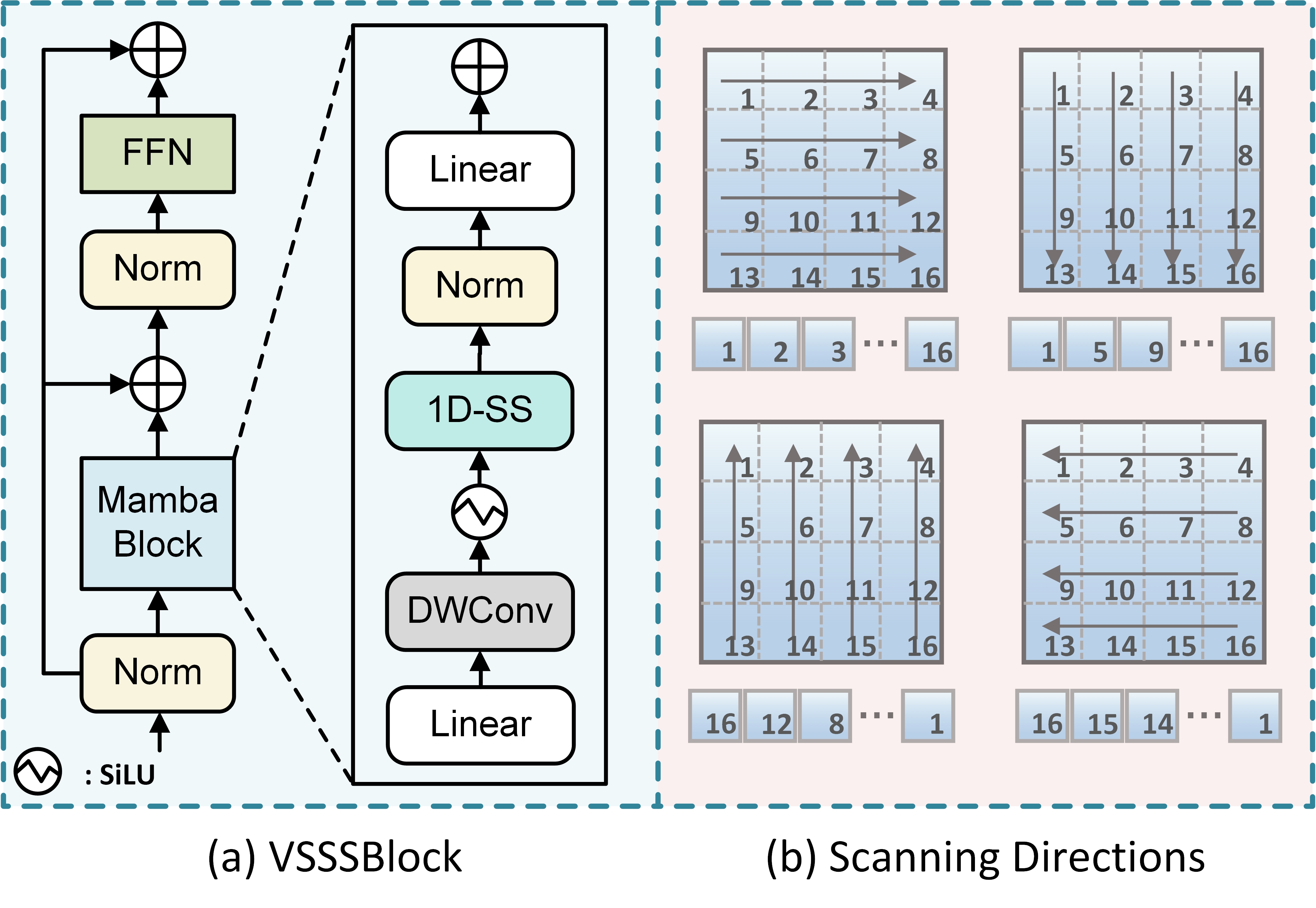} 
    \caption{(a) VSSSBlock. The 1D-SS refers to the use of a 1D selective scanning strategy, with the scanning direction of each VSSSBlock being one of the directions shown in the right. (b) Four different scanning directions.} 
    \label{fig:vsss} 
\end{figure}
While most time-domain TSE models typically use the convolution layers to extract speech embeddings, M3ANet adopts a more comprehensive approach by first capturing broad temporal features from the speech, followed by modeling long-range dependencies in each time window to extract deeper, more intricate speech characteristics. The proposed speech encoder, depicted in Figure~\ref{fig:overall}, comprises multi-scale encoding and GM modules~\cite{shaker2024groupmamba}.
Firstly, four Conv1D layers with different kernel sizes are applied to the mixed speech waveforms, each followed by a ReLU activation function, respectively, mapping the number of channels to $N_s$ to obtain richer speech information. Given mixture $X \in \mathbb{R}^{B \times 1 \times T}$, where $B$ represents the batch size, and $T$ is the length of the time-domain sequence, let $L_i$ represent the filter length at the $i$-th scale for $i \in \{1, 2, 3, 4\}$. The process is described as:
\begin{equation}
    X_i=ReLU(Conv1D(X,1,N_s,L_i))\in\mathbb{R}^{B \times N_s\times T_s}
\end{equation}
where $X_i$ denotes the encoded features at the $i$-th scale, and the stride of all parallel convolutions is set to $L_1 / 2$. By defining different $L_i$, these layers extract features at different temporal resolutions, capturing both short- and long-term dependencies within the speech signals.

To further extract the deeper information contained in speech with linear complexity, N layers of GM modules are incorporated after multi-scale encoding for long sequence modeling. The GM module divides the 2D input features into four groups along the channel dimension. Each group is independently processed by a visual single selective scanning block (VSSSBlock), which integrates a mamba block operating in a specific direction—left-to-right, right-to-left, top-to-bottom, or bottom-to-top—as illustrated in Figure~\ref{fig:vsss}. The full GM module also includes a channel affinity modulation (CAM) block and a feedforward layer.

M3ANet treats speech features at different granularities as channel-grouped 2D data. Specifically, $X_i$ is first expanded by adding a new dimension at the end and concatenated to form $\hat{X} \in \mathbb{R}^{B \times N_s\times T_s \times 4}$, which serves as the input to the GM block. Taking the first-layer GM block as an example, $\hat{X}$ is then split along the last dimension to recover each $X_i$, which are individually processed by the corresponding VSSSBlocks. The process can be described as follows:
\begin{equation}
    \resizebox{.91\linewidth}{!}{$
        \displaystyle
    X_i' = VSSSBlock(X_i,CrossScan_i)\in \mathbb{R}^{B \times N_s \times T_s \times 1}
    $}
\end{equation}
where $CrossScan_i$ represents the $i$-th scanning direction. Then, $X_i'$ is also concatenated along the last expanded dimension to obtain $X_c$. At this point, $X_c$ contains information from four scales and feature representations from four scanning perspectives.

Subsequently, the CAM block is used to enhance cross-scale feature interaction by reweighting the scale information. The structure of CAM can be represented as:
\begin{equation}
    X_{CAM}=X_c \otimes (\alpha W_2 \cdot \delta (W_1\hat{X}))
\end{equation}
where $\delta$ and $\alpha$ denote the ReLU and sigmoid functions, respectively, and $W_1$, $W_2$ are the learnable weights of fully connected layers. Next, $X_{\text{CAM}}$ is passed through a normalization layer and a feedforward layer, and then a residual connection is applied with the initial input $\hat{X}$, producing the output of the current GM module. To achieve optimal feature extraction, the GM module is repeated $N$ times. In summary, the speech encoder in this paper ultimately generates speech embeddings $\tilde{X} \in \mathbb{R}^{B \times N_s \times T_s \times 4}$.

\subsection{Modal Alignment Module}
Existing brain-assisted target speaker models focus on efficiently fusing EEG and speech information, yet they overlook the latency issue inherent in the data acquisition process, leading to misalignment of time frames. Inspired by the concept of ``align before fusion'' in~\cite{li2021align}, this paper proposes a multi-modal alignment framework based on contrast learning between speech and EEG embeddings, utilizing InfoNCE loss to optimize the alignment process. As shown in the red arrows of Figure~\ref{fig:overall}, the speech features $\tilde{X}$ are first reshaped to $(B \times N_s \times T_s \times 1)$ and then squeezed along the last dimension. Convolutional layers are applied to reconfigure the channels to match the EEG features $E'$ in dimension. To satisfy the computation requirements of InfoNCE, padding operation was applied to make $E'$ equal in temporal dimension with the speech sequence $\tilde{X}$, and the new vector obtained is $\tilde{E} \in \mathbb{R}^{B \times N_e \times T_s}$, and then the EEG and speech embeddings are unified to the shape $(B, N_e\times T_s)$.

The strategy for constructing sample pairs is also illustrated in Figure~\ref{fig:overall}. For the input EEG sequence $\{e_1, e_2, ..., e_n\}$ and speech sequence $\{x_1, x_2, ..., x_n\}$, where $n$ represents the batch size here, each EEG segment $e_i$ is treated as a query vector $q_i$. The corresponding speech frame $x_i$ in the time series is considered the positive sample, and both of them form a positive sample pair $(q_i, x_i^+)$. Meanwhile, other speech frames $x_j (j \neq i)$ from the same mini-batch are treated as negative samples, resulting in $n-1$ negative sample pairs $(q_i, x_j^-)$. To reduce time-level modality inconsistencies and achieve contrastive learning, we use the InfoNCE loss function for optimization, aiming to maximize the similarity between positive samples and minimize the similarity between negative samples. For each sample pair $(q_i, x_i^+)$, the InfoNCE function is defined as:
\begin{equation}
    \mathcal{L}_\mathrm{InfoNCE}=-\log\frac{\exp(sim(q_{i},x_i^{+})/\tau)}{\sum_{k=1}^n\exp(sim(q_{i},x_k)/\tau)}
\end{equation}
where $x_k$ represents speech segments in the current batch, sim(·, ·) represents the cosine similarity calculation between two samples, and $\tau$ is the temperature coefficient, set to 0.1.

Based on this contrastive learning strategy, the modal alignment module minimizes the distance between the EEG and the corresponding speech during training, achieving alignment between the two on a temporal frame level. This kind of cross-modal temporal alignment not only captures additional contextual information to enhance temporal consistency but also generates optimized multi-modal representations for downstream extraction tasks~\cite{ijcai2023p564}.

\subsection{Speaker Extractor}
The speaker extractor network, which serves as the core of the system, estimates a mask to extract the target speaker’s speech. To leverage cross-modal information, speech and EEG embeddings are fused prior to mask estimation. The proposed extractor comprises a modality fusion module implemented using convolutional multi-layer cross attention (CMCA)~\cite{Zhang2023basen}, and four layers of dual-path recurrent neural networks (DPRNN)~\cite{luo2020dual}.

CMCA is an efficient information fusion method. In this module, EEG and speech features are processed separately through information processing flows that include cross-attention, residual connection, and group normalization, enabling bidirectional feature coupling. CMCA employs a three-layer structure, where the output vector of the last layer is concatenated with the input vector of the first layer and processed with a 1D convolution to obtain the fused feature $Y \in \mathbb{R}^{B \times N_e \times T_s}$.

DPRNN, as a classical speech separation network, performs chunking processing on $Y$ with each chunk having a length of $L$ and a 50\% overlap, applying RNN operations in both intra-chunks and inter-chunks. Compared to CNNs, RNNs are more suitable for processing sequential data, as their recurrent structure effectively captures temporal dependencies. After processing through four layers of DPRNN, the network outputs the target speech mask $M \in \mathbb{R}^{B \times N_s \times T_s}$, which is element-wise multiplied with the encoded speech $X_{en}$ to obtain the masked speech $\hat{S}$. $X_{en} \in \mathbb{R}^{B \times N_e \times T_s}$ is obtained by concatenating the multi scale features $X_i$ along the channel dimension, followed by a convolution with $N_e$ output channels. The process is as follows:
\begin{equation}
    \hat{S}=X_{en}\otimes M \in \mathbb{R}^{B\times N_s\times T_s}
\end{equation}
Finally, the one-dimensional transposed convolution within the speech decoder reconstructs the estimated target speech $\hat{s}$ from $\hat{S}$.

\subsection{Objective Functions}
The objective function used in this study consists of two components: the SI-SDR loss function and the InfoNCE loss function. The SI-SDR loss is commonly employed in speech processing tasks and is defined as follows:
    \begin{equation}
    \mathcal{L}_{\mathrm{SI-SDR}}(s,\hat{s})=-10\log_{10}\frac{\left\|\frac{<\hat{s},s>}{\|s\|^2}s\right\|}{\left\|\hat{s}-\frac{<\hat{s},s>}{\|s\|^2}s\right\|}
    \end{equation}
where $s$ represents the ground truth speech and $\hat{s}$ represents the predicted target speaker speech extracted from the M3ANet. We combine the $\mathcal{L}_{\mathrm{SI-SDR}}$ and the contrastive learning loss $\mathcal L_{InfoNCE}$ to form the overall objective function, defined as follows:
    \begin{equation}
        \mathcal L_{total} = \mathcal L_{SI-SDR} + \lambda \mathcal L_{InfoNCE}
    \end{equation}
where $\lambda$ is the weighting factor and the value has been determined by experiment to be 3.

\section{Experimental Setup}
\subsection{Datasets}
\paragraph{Cocktail Party.}
For this dataset~\cite{broderick2018electrophysiological}, 33 subjects (aged 23-38 years) perform 30 trials, each lasting 60 seconds. During each trial, they listen to two classic works of fiction: one in the left ear and the other in the right ear. The subjects are divided into two groups, each instructed to focus on either the left or right ear. EEG data are recorded with 128 channels (plus two mastoid channels) at a rate of 512 Hz. For each subject, five trials are randomly selected as the test set, two for validation, and the remaining trials are used for training.

\paragraph{AVED.}
The audio-video EEG dataset (AVED)~\cite{zhang2024aved} includes EEG signals and corresponding speech from 20 normal-hearing students (14 males, 6 females) in a cocktail party scenario. Each completes 16 trials, and each trial lasts 152 seconds. EEG signals are recorded with 32 electrodes at a 1k Hz sampling rate. As in~\cite{fan2024msfnet}, the speaker-specific setting is applied to the AVED, analyzing only the trials where subjects focus on the left ear. For each subject, one trial is randomly selected as the test set, one as the validation set, and the remaining chosen trials are used for training.

\paragraph{MM-AAD.}
This dataset is a multi-modal AAD dataset~\cite{fan2025seeing}, consisting of EEG recordings from 50 normal-hearing subjects (34 males, 16 females). Each subject completes approximately 165 seconds of 20 trials. EEG data are captured from 32 channels at a 2048 Hz sampling rate. The attention direction used in AVED is also selected for MM-AAD. Similarly, for each subject, two trials are randomly selected as the test set, one as the validation set, and the remaining trials are used for training.

\subsection{Data Processing}
Raw EEG signals often contain a lot of noise, requiring preprocessing before they can be used. For the Cocktail Party dataset, EEG signals are downsampled to 128 Hz, filtered between 0.1-45 Hz, and re-referenced to the mastoid average. Independent component analysis (ICA) is then applied to remove artifacts from blinks and muscle movements. For the AVED and MM-AAD datasets, we apply a notch filter to remove 50 Hz frequency interference, followed by band-pass filtering (1-50 Hz) and downsampling the EEG signals to 128 Hz. ICA is used to separate mixed signal components and remove interference. Finally, EEG signals are re-referenced using a whole-brain average reference.

For all three datasets, the audio stimuli are downsampled to 14.7 kHz. EEG and speech signals are cut into 2-second segments for training and validation sets, while testing segments are 20 seconds long. No overlap occurs between data in different sets or segments.

\begin{table*}[t!]
    \centering
    \small
    \setlength{\tabcolsep}{10pt}
    \begin{tabular}{@{}ccccccc@{}}
    \toprule
    Datasets & Methods & SDR(dB) & SI-SDR(dB) & STOI & ESTOI & PESQ \\ \midrule
     & Mixture & 0.47 & 0.45 & 74.00\% & 55.00\% & 1.61 \\
     & BESD~\cite{hosseini2021speaker} & - & 5.75 & 79.00\% & - & 1.79 \\
     & UBESD~\cite{hosseini2022end} & - & 8.54 & 83.00\% & - & 1.97 \\
     & BASEN~\cite{Zhang2023basen} & - & 12.23 & 86.00\% & - & 2.24 \\
     & NeuroHeed*~\cite{pan2024neuroheed} & -0.09 & -0.11 & 71.48\% & 54.79\% & 1.45 \\
     & MSFNet~\cite{fan2024msfnet} & 13.03 & 12.89 & 88.00\% & 77.00\% & 2.51 \\
    \multirow{-7}{*}{Cocktail Party} & M3ANet (ours) & \textbf{14.11} & \textbf{13.95} & \textbf{89.23\%} & \textbf{78.36\%} & \textbf{2.58} \\ \midrule
     & Mixture & 1.54 & 1.52 & 75.83\% & 60.57\% & 1.50 \\
     & UBESD~\cite{hosseini2022end} & 8.1 & 7.89 & 85.00\% & 72.00\% & 1.75 \\
     & BASEN~\cite{Zhang2023basen} & 8.68 & 8.46 & 86.00\% & 75.00\% & 1.91 \\
     & NeuroHeed*~\cite{pan2024neuroheed} & 8.77 & 8.61 & 88.11\% & 77.81\% & 1.82 \\
     & MSFNet~\cite{fan2024msfnet} & 9.84 & 9.65 & 89.00\% & 79.00\% & 2.07 \\
    \multirow{-7}{*}{AVED} & M3ANet (ours) & \textbf{11.14} & \textbf{10.89} & \textbf{90.60\%} & \textbf{82.06\%} & \textbf{2.21} \\ \midrule
     & Mixture & 2.14 & 2.12 & 79.67\% & 64.10\% & 1.54 \\
     & BESD*~\cite{hosseini2021speaker} & 6.46 & 5.44 & 83.94\% & 66.09\% & 1.68 \\
     & UBESD*~\cite{hosseini2022end} & 6.60 & 6.29 & 84.32\% & 69.59\% & 1.66 \\
     & BASEN*~\cite{Zhang2023basen} & 8.22 & 8.04 & 87.72\% & 74.45\% & 2.01 \\
     & NeuroHeed*~\cite{pan2024neuroheed} & 8.42 & 8.26 & 89.10\% & 78.10\% & 2.18 \\
     & MSFNet*~\cite{fan2024msfnet} & 9.69 & 9.57 & 90.19\% & 78.71\% & 2.18 \\
    \multirow{-7}{*}{MM-AAD} & M3ANet (ours) & \textbf{10.71} & \textbf{10.57} & \textbf{91.68\%} & \textbf{82.31\%} & \textbf{2.33} \\ \bottomrule
    \end{tabular}
    \caption{Comparison with other mainstream models on Cocktail Party, AVED, MM-AAD. The experimental results marked with an asterisk (*) were reproduced using the authors' publicly available code. Here, NeuroHeed operates in a non-causal setting. For the Cocktail Party dataset, the other experimental results are obtained from their respective papers. In contrast, the results on the AVED dataset are derived from \protect\cite{fan2024msfnet}.}
    \label{table:data}
\end{table*}

\subsection{Training Details}
In the training process, the model was trained for 60 epochs using the Adam optimizer with a weight decay of $1 \times 10^{-3}$ and momentum of 0.9. The learning rate scheduler used in this study combines a linear warm-up phase (with a warm-up ratio of $5\%$) followed by a cosine decay phase. The learning rates are tailored for each dataset to suit them better. Specifically, the initial learning rate is set to $6e^{-4}$ for the Cocktail Party dataset, $2e^{-3}$ for AVED and $3e^{-3}$ for MM-AAD. The batch size was set to 8, and all experiments were accelerated by a single NVIDIA GeForce RTX 4090 GPU.

For our network, we set the speech feature dimension $N_s$ to 128 and EEG feature dimension $N_e$ to 64. The speech encoder uses filter lengths of 2.5ms, 5ms, 10ms, and 20ms ($L_1$ to $L_4$). In the DPRNN, the size $L$ of each block is set to 250.

\subsection{Evaluation Metrics}
We focus on using five objective evaluation metrics to measure the quality of predicted speech. The source to distortion ratio (SDR)~\cite{le2019sdr} and scale-invariant source to distortion ratio (SI-SDR) measure the quality of a signal, with SI-SDR being more robust to scaling differences. Short-time objective intelligibility (STOI) and extended short-time objective intelligibility (ESTOI)~\cite{taal2011algorithm} assess speech intelligibility, with ESTOI offering more accurate predictions by considering nonlinear hearing properties, with values ranging from 0 to 1. Perceptual evaluation of speech quality (PESQ)~\cite{rix2001perceptual} compares the loudness spectra of desired and separated speech signals to evaluate speech quality. The PESQ scores range from $\left [ -0.5, 4.5 \right ] $. These metrics are widely used in speech applications due to their objective nature and ability to accurately reflect signal quality, intelligibility, and perceptual performance. Higher values are better for all metrics. Among them, SDR and SI-SDR are measured in dB, while the other metrics are unitless.

\section{Results}
\subsection{Comparison with Baseline Methods}
We compare M3ANet with current brain-assisted target speaker extraction models, including BESD, UBESD, BASEN, NeuroHeed, and MSFNet. The experimental results on three datasets, presented as the median of all test segments from the subjects, are summarized in Table~\ref{table:data}. As shown, our M3ANet outperforms all other models across all three datasets.

On the Cocktail Party dataset, M3ANet outperforms the current SOTA model, MSFNet, with significant improvements across all metrics. Specifically, M3ANet achieves improvements of 1.08 dB in SDR, 1.06 dB in SI-SDR, 1.23\% in STOI, 1.36\% in ESTOI, and 0.07 in PESQ. Moreover, with an SDR of 14.11 dB and an SI-SDR of 13.95 dB, the results demonstrate that our proposed model achieves competitive performance in speaker extraction.

On the AVED dataset, M3ANet also shows notable improvements over MSFNet. It achieves 1.3 dB in SDR, 1.24 dB in SI-SDR, 1.6\% in STOI, 3.06\% in ESTOI, and 0.14 in PESQ improvements. Notably, the significant improvements in SDR and SI-SDR highlight M3ANet's effectiveness in enhancing speech quality and reducing distortion. Similarly, on the MM-AAD dataset, M3ANet delivers impressive performance, surpassing the current best performance by 1.02 dB in SDR, 1 dB in SI-SDR, 1.49\% in STOI, 3.6\% in ESTOI, and 0.15 in PESQ, which further demonstrates the outstanding generalizability and efficiency of the proposed model. The notable ESTOI gains on both datasets indicate that M3ANet also effectively improves speech intelligibility.

\begin{figure}[] 
    \centering
    \includegraphics[width=0.4\textwidth]{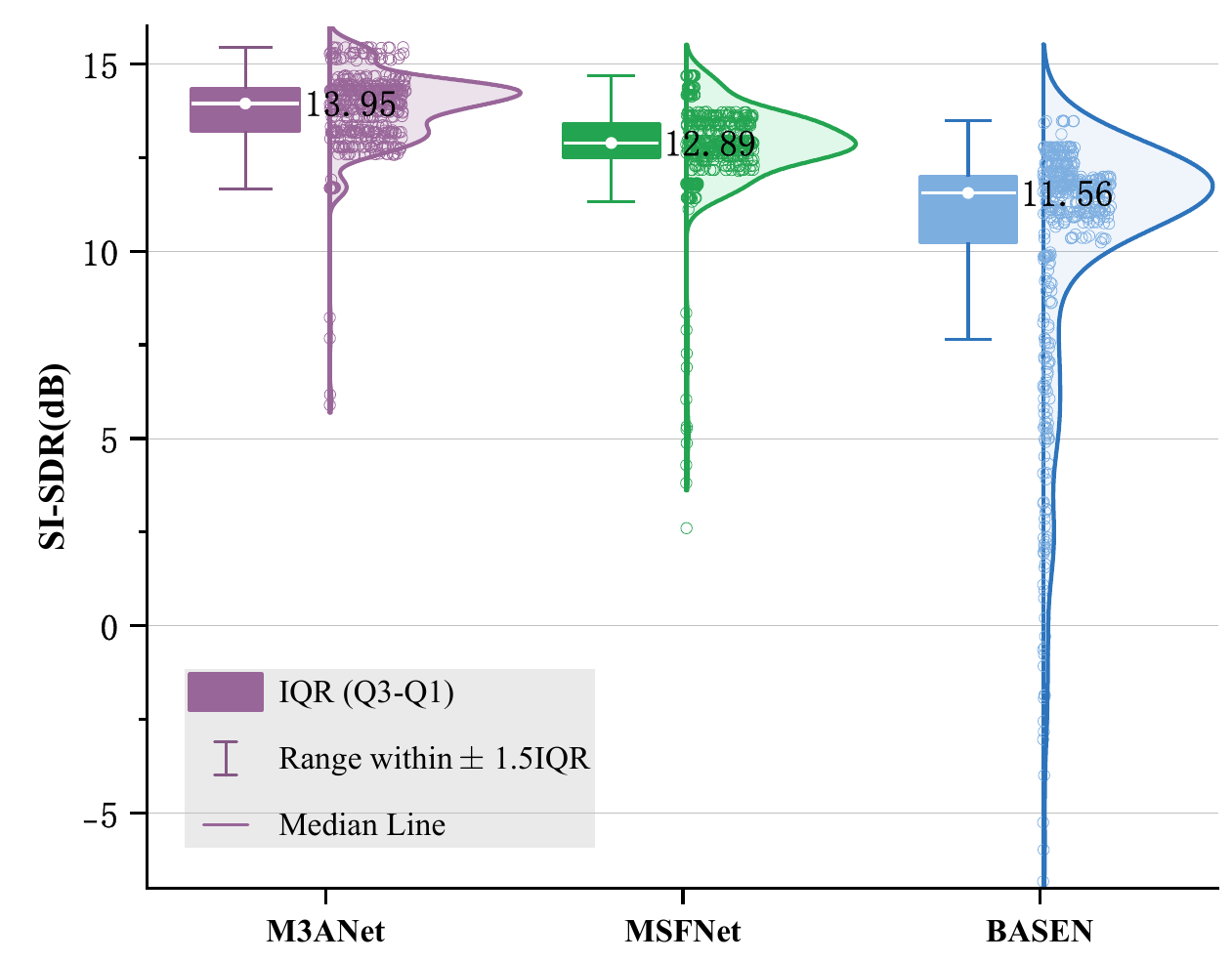} 
    \caption{Combined boxplot and half-violin plot comparison of SI-SDR (dB) performance across different models. The colored circles represent the SI-SDR values for each test speech segment. The right half-violin plot indicates the density and distribution of the data, where a wider area indicates a higher concentration of data points. The left boxplot includes the median line, interquartile range (IQR), and whiskers, which represent the range within $\pm$ 1.5 IQR. Values outside this range are considered outliers. } 
    \label{fig:scatter} 
\end{figure}

Additionally, we present the combined boxplot and half-violin plot for the extracted speech segments from BASEN, MSFNet, and M3ANet models on the Cocktail Party dataset, as shown in Figure~\ref{fig:scatter}. The experimental results of all models are obtained from our own reproductions. The SI-SDR results of BASEN are widely distributed, accompanied by high variance and many low or even negative values, suggesting unstable performance and weak generalization. MSFNet shows more stability, with a narrower distribution and fewer outliers. In contrast, M3ANet outperforms both BASEN and MSFNet, with a concentrated distribution around its high median value of 13.95 dB and almost no outliers, highlighting its superior and stable performance.

These results underscore the importance of maintaining temporal consistency between EEG and speech signals, while also highlighting the significance of fully extracting features from mixed speech. Furthermore, the results highlight the model's potential for real-world applications in brain-assisted speech extraction, offering a promising solution to the challenges posed by multi-modal data integration.

\subsection{Ablation Study}

\begin{table}[t!]
    \centering
    \small
    \setlength{\tabcolsep}{3pt} 
    \begin{tabular}{@{}ccccc@{}}
    \toprule
    Datasets & Methods & SDR(dB) & SI-SDR(dB) & STOI \\ \midrule
     & M3ANet (ours) & \textbf{14.11} & \textbf{13.95} & \textbf{89.23\%} \\
     & w/o GM & 13.95 & 13.80 & 88.94\% \\
    \multirow{-3}{*}{Cocktail Party} & w/o Alignment & 13.78 & 13.63 & 88.87\% \\ \midrule
     & M3ANet (ours) & \textbf{11.14} & \textbf{10.89} & \textbf{90.60\%} \\
     & w/o GM & 10.17 & 9.96 & 88.93\% \\
    \multirow{-3}{*}{AVED} & w/o Alignment & 10.25 & 10.04 & 88.88\% \\ \midrule
     & M3ANet (ours) & \textbf{10.71} & \textbf{10.57} & \textbf{91.68\%} \\
     & w/o GM & 10.04 & 9.87 & 90.25\% \\
    \multirow{-3}{*}{MM-AAD} & w/o Alignment & 10.12 & 9.99 & 91.43\% \\ \bottomrule
    \end{tabular}
    \caption{Ablation study on three dataset. \textbf{w/o GM} stands for removing the GM block from the speech encoder and retaining only the multi-scale encoding structure. \textbf{w/o Alignment} means removing the modal alignment module between EEG and speech embeddings.}
    \label{table:ablation}
\end{table}

To investigate the effectiveness of the GM layer and modal alignment module, we conducted ablation experiments on all three datasets, with the results shown in Table~\ref{table:ablation}. We primarily present three key speech quality evaluation metrics: SDR, SI-SDR and STOI. 

\paragraph{Effectiveness with GM Block.}In the w/o GM model, we removed the GM module from the speech encoder but still used the four-scale Conv1D for initial feature extraction. To be consistent with the structure of the proposed model, we use a simple addition operation to fuse features of different granularities and input them into the speaker extractor. The experimental results demonstrate that removing the GM module led to a decrease in SDR, SI-SDR, and STOI across all three datasets, especially in the AVED and MM-AAD datasets, where SDR values decreased by 0.97 dB and 0.67 dB, SI-SDR values dropped by 0.93 dB and 0.7 dB, and STOI declined by 1.67\% and 1.43\%, respectively. These findings not only highlight the critical role of the GM module in maintaining model performance but also demonstrate its profound capability to extract latent speech features in depth. Thanks to the powerful long sequence modeling capability of the mamba module, the GM module can extract the speech embeddings more adequately and efficiently. Meanwhile, the CAM structure within the GM module acilitates information interaction across temporal scales, thereby enhancing both separation performance and speech extraction quality.

\paragraph{Effectiveness with Alignment Module.}To explore the impact of the modal alignment module on overall performance, we removed it between EEG and speech embeddings, while keeping the rest of the model architecture consistent with M3ANet. Compared to the final results, removing the alignment module led to a decrease in all performance metrics. On the three datasets, the SDR values dropped by 0.33 dB, 0.89 dB, and 0.59 dB, respectively, and the SI-SDR values decreased by 0.32 dB, 0.85 dB, and 0.58 dB. Moreover, the STOI scores show varying degrees of degradation across different datasets. These results highlight the critical role of the alignment module in enhancing model performance for target speaker extraction. The performance degradation without this module can be attributed to the lack of feature alignment between the EEG and speech modalities. By optimizing the InfoNCE loss, the module effectively reduces the distance between positive pairs and aligns features at corresponding time steps across modalities, thereby improving separation quality through more accurate temporal correspondence between EEG and speech signals.

\subsection{Impact of GM Layers on Model Performance}
Additionally, we investigated the impact of the number of GM layers on overall performance. As shown in Table~\ref{table:layers}, we explored the model results with 1 to 5 GM layers. The results indicate that the model with 2 layers of GM achieved the best performance across SDR, SI-SDR, and STOI. As the number of GM layers increased, the overall performance of the model deteriorated. This experiment demonstrates that the GM module is a simple and efficient feature extraction method, which can significantly improve model performance with only a small increase in model complexity while maintaining good training efficiency and generalization ability. 
\begin{table}[]
    \small
    \centering
    \setlength{\tabcolsep}{4pt} 
    \begin{tabular}{@{}cccccc@{}}
    \toprule
    Layers & SDR(dB) & SI-SDR(dB) & STOI & ESTOI & PESQ \\ \midrule
    1 & 13.25 & 13.08 & 87.96\% & 75.08\% & 2.48 \\
    2 & \textbf{14.11} & \textbf{13.95} & \textbf{89.23\%} & 78.36\% & 2.58 \\
    3 & 13.97 & 13.83 & 88.91\% & \textbf{78.42}\% & \textbf{2.61} \\
    4 & 13.71 & 13.55 & 89.20\% & 77.90\% & 2.58 \\
    5 & 13.09 & 12.89 & 87.93\% & 75.97\% & 2.54 \\ \bottomrule
    \end{tabular}
    \caption{Impact of the number of layers of GM on overall performance. Experiments are conducted on the Cocktail Party dataset.}
    \label{table:layers}
\end{table}

\section{Conclusion}

This paper proposes a novel time-domain, multi-scale and multi-modal alignment network for brain-assisted target speaker extraction. To align EEG and speech temporally, we introduce a modal alignment module based on contrastive learning, which reduces temporal discrepancies between the two embeddings. Additionally, to fully capture speech features, we employ multi-scale encoding to extract broad temporal features and use GM modules to extract more detailed and deeper speech information. Experimental results on three datasets demonstrate that the proposed model achieves state-of-the-art performance and strong generalization capabilities. In future work, we will explore advanced contrastive learning strategies and more efficient feature fusion techniques to further enhance the alignment and complementarity between the two modalities.

\section*{Acknowledgments}

This work is supported by the {STI 2030—Major Projects (No. 2021ZD0201500)}, the National Natural Science Foundation of China (NSFC) (No.62201002, 6247077204), Excellent Youth Foundation of Anhui Scientific Committee (No. 2408085Y034), Distinguished Youth Foundation of Anhui Scientific Committee (No. 2208085J05), Special Fund for Key Program of Science and Technology of Anhui Province (No. 202203a07020008), Cloud Ginger XR-1.

\bibliographystyle{named}
\bibliography{reference}

\end{document}